# Datasets of ionospheric parameters provided by SCINDA GNSS receiver from Lisbon airport area

Tatiana Barlyaeva*, Teresa Barata, Anna Morozova

*Univ Coimbra, Center for Earth and Space Research of the University of Coimbra, Almas de Freire, Sta. Clara, 3040-004, Coimbra, Portugal*
*SWAIR project*

*TVBarlyaeva@gmail.com

**Abstract**

Here we present datasets provided by a SCINDA GNSS receiver installed in the Lisbon airport area from November of 2014 to July of 2019. The installed equipment is a NovAtel EURO4 with a JAVAD Choke-Ring antenna. The data are in an archived format and include the general messages on quality of records (*.msg), RANGE files (*.rng), raw observables as the signal-to-noise (S/N) ratios, pseudoranges and phases (*.obs), receiver position information (*.psn), ionosphere scintillations monitor (ISMRB; *.ism) and ionospheric parameters: total electron content (TEC), rate of change of TEC index (ROTI), and the scintillation index S4 (*.scn). The presented data cover the full 2015 year. The raw data are of 1-minute resolution and available for each of the receiver-satellite pairs. The processing and the analysis of the ionosphere scintillation datasets can be done using a specific "SCINDA-Iono" toolbox for the MATLAB developed by T. Barlyaeva (2019) and available online via MathWorks File Exchange system. The toolbox calculates 1-hour means for ionospheric parameters for each of the available receiver-satellite pairs and averaged over all available satellites during the analyzed hour. Here we present the processed data for the following months in 2015: March, June, October, and December. The months were selected as containing most significant geomagnetic events of 2015. The 1-hour means for other months can be obtained from the raw data using the aforementioned toolbox. The provided datasets are interesting for the GNSS and ionosphere based scientific communities.

**Specifications Table**

| Subject | Signal Processing |
|---|---|
| Specific subject area | The datasets provided by GNSS receivers, in particular ionosphere scintillation data. |
| Type of data | ASCII, text messages and binary<br>Table |
| How data were acquired | The data have been recorded by a GNSS receiver installed in Lisbon airport area.<br>Instruments:the installed equipment is a NovAtel EURO4 with a JAVAD Choke-Ring antenna and a firmware (SCINDA) installed.<br>SCINDA (Scintillation Network Decision Aid) is a system designed to specify ionospheric scintillation in real time and was developed by the US Air Forces. |
| Data format | Raw<br>Processed |
| Parameters for data collection | The data were acquired continuously at 1Hz and stored in a computer and external disks. In the Lisbon airport the electric power and limited access to people were ensured in order to maintain the quality and integrity of data. The equipment was not connected to the internet. System maintenance operations were performed periodically so that there was no interruption of data collection. No calibration of the installed receiver was done. |
| Description of data collection | Data presented here were collected from January to December of 2015.The data were acquired continuously at 1Hz. The firmware SCINDA give a file as output with calculated receiver position along with time-tag and the number of satellites in view, including satellite identifier number (PRN) for the tracked ones. The acquired TEC data are without correction for the receiver bias. |

| **Data source location** | Institution: Lisbon airport area |
| --- | --- |
| | City/Town/Region: Lisbon |
| | Country: Portugal |
| | Latitude and longitude (and GPS coordinates) for collected samples/data: |
| | Latitude (degrees): 38.77932527 |
| | Longitude (degrees): -9.139699966 |
| | Altitude (ellipsoidal, m): 128.139 |
| **Data accessibility** | Repository name: Mendeley Data |
| | Data identification number: 10.17632/kkytn5d8yc.1 |
| | Direct URL to data: |
| | https://data.mendeley.com/datasets/kkytn5d8yc/1 |

**Value of the Data**

- Here we provide complementary data set allowing to estimate ionospheric conditions and GNSS positioning quality in 2015 for the mid-latitudinal region at the west coast of the Iberian Peninsula (area of the Lisbon airport).

- The data set contains both the raw and processed data from a GNSS receiver. The processing includes, e.g., gaps and bad data removal, averaging over 1-hour time interval, both for individual receiver-satellite pairs and over all available pairs. We also provide a tool to process and analyze the raw data: the "SCINDA-Iono" toolbox for the MATLAB, which is publicly available.

- These data can be used for the analysis of space weather effects on the GNSS signal quality and estimations of corrections to be used in upcoming GNSS services/systems accounting for the space weather conditions. Also, this data can be used for an analysis of ionosphere conditions observed in 2015.

- GNSS is widely used now in numerous applications for roads, railways, maritime, agriculture and air navigation, so the improvements of (i.e.) position information provided by GNSS services can serve for the safety of life.

- There are at least two main communities who can directly benefit from these data: the communities of ionosphere sciences and of the GNSS technologies and services.

- Some scientific results of the ionosphere conditions analysis during few significant geomagnetic storms in 2015 can be found in [1].

**Data Description**

Here we present the data for the full 2015 year provided by a SCINDA GNSS receiver installed in Lisbon airport area from November of 2014 to July of 2019. The installed equipment is a NovAtel EURO4 with a JAVAD Choke-Ring antenna, with the firmware Scintillation Network Decision Aid (SCINDA). The details of the SCINDA software functionalities can be found in [2-8].

The raw data (subfolder "DATA-RAW\") include the general messages on the quality of records (*.msg), RANGE files (*.rng), raw observables as S/N ratios, pseudoranges and phases (*.obs), receiver position information (*.psn), ionosphere scintillations monitor (*.ism) and ionosphere scintillations (*.scn) as TEC (Total Electron Content), S4 (Scintillation index) and ROTI (Rate ofTEC Index).

The folders with raw data are organized as follows (see Fig. 1). The folder for the year with the name in the format "YYYY\" (2015) contains subfolders entitled "MM-Mon\" that include sets of gz-archived files with the names in format "YYMMDD_HH0000" (where '0000' is the suffix that corresponds to the hour 00:00) and extensions ".ism", ".msg", ".rng", ".obs", ".psn" and ".scn". The ".msg" files contain text messages on the receiver diagnostics and records' quality, the ".ism" files keep alternative ionosphere monitor statistics (ISMRB) records written in verbatim from the receiver without using SCINDA software, and the ".rng" files are RANGE records. The ".obs", ".psn" and ".scn" files contain data in the ASCII format. Each of the latter data files covers 1-hour interval and consists of the data of 1-minute resolution for all visible satellites. Each data sequence in these files starts from the epoch information (date and time) and followed by the data. The ".obs" files present raw records on the S/N ratios, pseudoranges and phases for both the L1 and L2 channels. The ".psn" files contain latitude, longitude and altitude followed by the used satellites information. The ".scn" files report the detailed ionosphere conditions' information (see for ".scn" records section on the processed ".scn" data in the "Experimental Design, Materials, and Methods" with Figure 8).

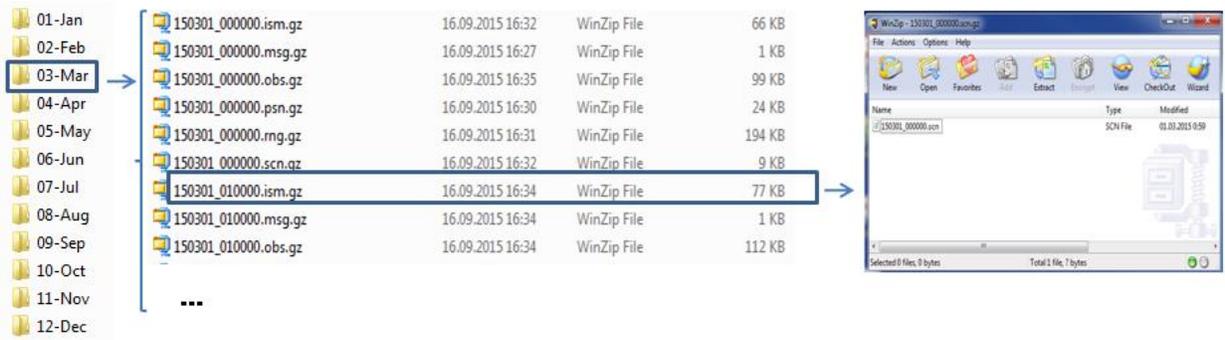

Figure 1. Organization of the raw data in subfolders of the folder "2015\".

In addition to the raw data, kept in the archived format, we present the processed scintillation (*.scn) datasets for selected months of 2015: March, June, October and December, the months with most significant geomagnetic events of 2015. These data can be found in the subfolder "DATA-PROCESSED\". One should note that in theory SCINDA software is able to do separation of the ionosphere and plasmasphere TEC [3], but for the presented data this option was deactivated.

The processed scintillation data are organized in the following way (see Fig. 2). For example, the data in the folders "All_T20_61p_TwD_2015-03\" and "SATs_T20_61p_TwD_2015-03\" are the March 2015 (see "2015-03") data processed to eliminate three types of errors that can be found in the raw data files, which we named as "T20", "61p" and "TwD" errors. The meaning and details of these types of erroneous or missed blocks are explained below in "Experimental Design, Materials, and Methods" section, "Software functionalities" subsection. The 'T20_61p_TwD' labels in the folders' names mean that the processed data are free from errors of all three types. These folders contain either data averaged over all visible satellites data ('All' case) or the data for each of receiver-satellite pairs separately ('SATs' case). For the case of the separate receiver-satellite pairs the list of the visible satellites is provided in a dedicated file. Each folder contains pairs of files: the data file and the file with time stamps (1-minute resolution).

| Name | Date modified | Type | Size |
|---|---|---|---|
| All_T20_61p_TwD_2015-03 | 20.12.2019 13:15 | File folder | |
| All_T20_61p_TwD_Means1h_2015-03 | 20.12.2019 13:15 | File folder | |
| SATs_T20_61p_TwD_2015-03 | 20.12.2019 13:15 | File folder | |
| SATs_T20_61p_TwD_SATs1h_2015-03 | 20.12.2019 13:15 | File folder | |

Figure 2. Organization of the processed data in corresponding folders.

The subfolders "All_T20_61p_TwD_Means1h_2015-03\" and "SATs_T20_61p_TwD_SATs1h_2015-03\" contain the 1-hour means for the averages over all available satellites and for each of the receiver-satellite pairs, respectively. Each 1-hour means file also contains the standard deviation and the number of successful observations. The list of visible satellites is also available in a separate file.

| Name | Date modified | Type | Size |
| --- | --- | --- | --- |
| SCN_2015-03_res1m_01-31.dat | 20.12.2019 12:41 | DAT File | 4 129 KB |
| SCN_2015-03_res1m_01-31_DATES-TIMES.dat | 20.12.2019 12:41 | DAT File | 4 129 KB |

Figure 3. Organization of the processed data files: 1-minute data for averages over all visible satellites.

| Name | Date modified | Type | Size |
| --- | --- | --- | --- |
| SATs_LIST_UNIQUES_201503.dat | 20.12.2019 11:56 | DAT File | 1 KB |
| SCN_2015-03_res1m_01-31_1.dat | 20.12.2019 12:33 | DAT File | 1 192 KB |
| SCN_2015-03_res1m_01-31_1_DATES-TIMES.dat | 20.12.2019 12:33 | DAT File | 1 192 KB |
| SCN_2015-03_res1m_01-31_2.dat | 20.12.2019 12:33 | DAT File | 1 269 KB |
| SCN_2015-03_res1m_01-31_2_DATES-TIMES.dat | 20.12.2019 12:33 | DAT File | 1 269 KB |
| SCN_2015-03_res1m_01-31_3.dat | 20.12.2019 12:33 | DAT File | 1 055 KB |
| SCN_2015-03_res1m_01-31_3_DATES-TIMES.dat | 20.12.2019 12:33 | DAT File | 1 055 KB |

Figure 4. Organization of the processed data files: 1-minute data for each receiver-satellite pair.

| Name | Date modified | Type | Size |
| --- | --- | --- | --- |
| SCN_2015-03_Means1h_01-31.dat | 20.12.2019 12:41 | DAT File | 69 KB |
| SCN_2015-03_Means1h_01-31_DATES-TIMES.dat | 20.12.2019 12:41 | DAT File | 69 KB |
| SCN_2015-03_Means1h_01-31_Nobs.dat | 20.12.2019 12:41 | DAT File | 69 KB |
| SCN_2015-03_Means1h_01-31_Std.dat | 20.12.2019 12:41 | DAT File | 69 KB |

Figure 5. Organization of the processed data files: 1-hour means for averages over all visible satellites.

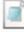

Figure 6. Organization of the processed data files: 1-hour data for each receiver-satellite pair.

Figures 3 to 6 illustrate how the processed data are organized in files. Examples of the data files for 1-minute resolution are in Figures 3 and 4 for the 'All' and 'SATs' cases, respectively. Similar data for the 1-hour resolution are in Figures 5 and 6.

The files with the processed data contain six columns that correspond, respectively from the 1st to the 6th, to 'L1S4', 'L2S4', 'TECP', 'TECF', 'ROTI', 'TECR' values which are the scintillation intensity index S4 for L1 and L2 bands, differential pseudorange TEC (in TEC Units, TECU), differential carrier phase TEC (in TECU), ROTI (rate of change of TEC over one minute), relative and not-calibrated TEC (in TECU), respectively. The files with dates/times contain six columns that correspond (from the 1st to the 6th), to the year, month, day, hour, seconds from the start of day, and time in parts of a day, respectively.

The uploaded data are accompanied by a short README file.

**Experimental Design, Materials, and Methods**

To process and analyze the raw ionospheric data we recommend to use the "SCINDA-Iono" toolbox for the MATLAB specifically developed for this purpose. This toolbox available online via the MathWorks FileExchange system and can be downloaded at *https://www.mathworks.com/matlabcentral/fileexchange/71784-scinda-iono_toolbox* .

There are other packages developed to analyze GNSS scintillation data provided by various receivers including SCINDA firmware, but some of them are proprietary, at least partly (e.g., GAMIT/GLOBIK/TRACK, http://www-gpsg.mit.edu/~simon/gtgk/, GAMIT is developed for Unix only), some are developed in other languages (e.g., GPSTk Post-processing, www.gpstk.com, in C++), and some are oriented mostly to position error and protection level calculation (e.g., RTKLIB, http://www.rtklib.com/). Some packages (e.g., "GPS-TEC" [6]) being developed for other environment can be run also in the MATLAB but do different kind of data processing and analysis comparing to our package.

The advantage of the "SCINDA-Iono" toolbox comparing to other packages for the analysis of the GNSS scintillation data from the SCINDA receivers is that it is free and publicly available, developed specifically for the MATLAB, can be run under Windows system, and is oriented to the general scintillation data preprocessing, processing and analysis. The toolbox is designed to analyze the data in the time interval chosen inside one calendar month.

*Software Architecture.*
The architecture of the "SCINDA-Iono" toolbox is schematically presented in Fig. 7. It can be formally divided into three principal blocks:
 - an adaptation of the receiver provided scintillation data for the further analysis;
 - their preprocessing;
 - analysis.

The preprocessing includes the corrections of the data for the receiver's miscalculations and the internal clock failures. There are three types of the preprocessing procedures (see Fig. 1).

*Software Functionalities.*
The "SCINDA-Iono" toolbox allows to perform following operations:
 - *To unzip the scintillation data.* Since the original data are provided in the archived format, to be processed they should be unzipped.
 - *To create the data files for each of the receiver-satellite pairs*, that have the format similar to one of the original scintillation data files.
 - *To delete epochs with erroneous values (so-called 'T20' preprocessing).* Some epochs start, as normally, by the letter 'T' and followed by '-20' (instead of a two-digit year number). Even if the content of such epochs seems to contain true observations we prefer to delete such epochs with erroneous headers.

 - *To combine all data for a specific hour from different files to a single file (so-called '61p' preprocessing).* Some hourly files of 1-minute data contain one measurement from the 'neighbor' hour, thus the 61th measurement appears. Such `extra' measurements are moved to the corresponding hour file.
 - *To delete the epochs without data, that contains only titles (so-called 'TwD' preprocessing).* Some blocks are empty and contain no data except the header.

- *To analyze the 1-minute resolution data.* As a result, the graphs of all preprocessed data with 1-minute time resolution are plotted.
 - *To analyze 1-hour resolution data.* Here the 1-hour means are calculated accompanied by the standard deviation (Std) and a number of successful observations (Numb. of Points).

It should be noted that 1) the data obtained on each step listed above are saved into indicated folders and 2) each of the steps listed above can be included or excluded by user upon decision. In the latter case, one must pay attention to the names of the files to be used for further steps.

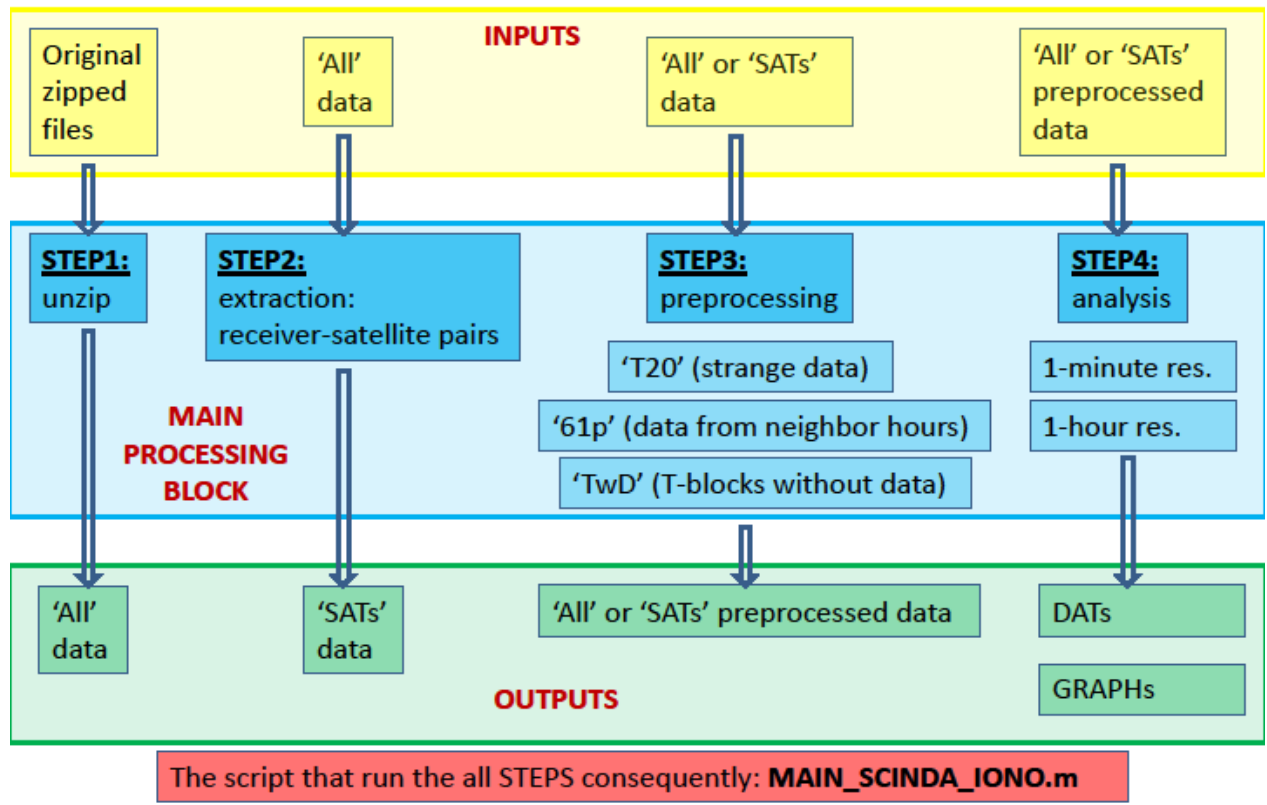

Figure 7: "SCINDA-Iono" toolbox architecture.

*Illustrative Examples.*
Here we present an example of the data observed during the most significant geomagnetic storm of the 24$^{th}$ solar cycle that took place on 17-18 of March, 2015. The original scintillation data (*.scn) are of the format presented in Fig. 8. This data file is an input file for the "SCINDA-Iono" toolbox.

```
        MM     UTSEC
   YY    DD
   ↓     ↓      ↓
   ↓     ↓      ↓
T 15 03 01 00032
323.3 58.7 0.03 100 0.02 100  18.3  -32.622  10.28 14.5 133  16
118.1 22.7 0.08 100 0.15 100  26.3  -12.477   8.15 28.7  38  18
261.7 16.3 0.14 100 0.17 100  30.6  -11.866   9.10 26.3  19  19
 61.3 55.6 0.05 100 0.04 100   9.8 -118.543   8.01 11.5 192  21
153.4 11.3 0.20 100 0.32 100  41.1   -4.568  10.99 40.8   4  22
279.6 43.0 0.07 100 0.03 100  42.1  -52.187   9.44 44.7  96  27
186.7 26.9 0.06 100 0.09 100  18.4  -31.935   7.43 39.2 297  31
190.1 44.8 0.05 100 0.00   0   0.0   -0.000   0.00  0.0   0 120
132.4 32.9 0.11 100 0.00   0   0.0   -0.000   0.00  0.0   0 126
T 15 03 01 00092
323.6 59.2 0.04 100 0.02 100  16.8  -32.659   8.87 14.5 134  16
117.7 22.9 0.13 100 0.16 100  31.2  -12.718  10.39 28.6  39  18
262.0 16.6 0.16 100 0.17 100  27.5  -12.409  10.80 25.8  20  19
  ↑    ↑    ↑        ↑         ↑      ↑        ↑    ↑    ↑   ↑
  AZ       L1S4      L2S4      TECP           ROTI       N
       EL       %SAM      %SAM       TECF          TECR      PRN
```

Figure 8: Example of an original scintillation (*.scn) data file.

Files of this type contain both satellite characteristics and ionospheric parameters. Each file consists of values registered by various satellites, and each line of the file contains data obtained by one receiver-satellite pair. Each epoch starts from the letter 'T' and contains year (YY, two digits), month (MM, two digits), day (DD, two digits), time (UTSEC, in seconds since midnight). The data of each epoch contains the azimuth of a satellite (AZ, in degrees), elevation of a satellite (EL, in degrees), scintillation intensity index S4 on the L1 and L2 bands (L1S4 and L2S4, respectively), per cent of the samples taken compared to number expected (%SAM, between 0 and 100%), differential pseudorange (TECP, in TECU), differential carrier phase (TECF, in TECU), relative not-calibrated TEC (TECR, in TECU), rate of change of TEC over one minute (ROTI), time since last time slip (N, in minutes) and the pseudorandom noise satellite identifier (PRN). As it was already mentioned, the receiver was not calibrated for the receiver bias.

The toolbox is developed to work with the original data of 1-minute resolution. As was already mentioned above, the software can do an extraction of the data for each of available receiver-satellite pairs. An example of the resulting data file for this case is presented in Fig. 9. The averages over all available for the moment satellites can be calculated as well. The corresponding 1-minute resolution plots for the average over all available satellites are presented in Fig. 10.

In addition, 1-hour means for each of the receiver-satellite pairs or average values over all available satellites are calculated. The resulting set of 1-hour resolution plots for the average over all available satellites is presented in Fig. 11. For the averages the main figure is accompanied by the plots of standard deviation (Fig. 12) and the number of successful observations (Fig. 13).The total time of the data processing and analysis takes, in general, not more than few minutes with the MATLAB R2018b on the Dell laptop with processor Intel(R) Core(TM) i5-3337U CPU @ 1.80GHz; installed 8,00 GB of RAM; with the 64-bit Operating system (Windows 7 Professional edition).

```
T 15 03 01 00032
323.3 58.7 0.03 100 0.02 100   18.3 -32.622   10.28   14.5 133 16
T 15 03 01 00092
323.6 59.2 0.04 100 0.02 100   16.8 -32.659    8.87   14.5 134 16
T 15 03 01 00152
323.8 59.6 0.03 100 0.03 100   13.8 -32.703    8.57   14.4 135 16
T 15 03 01 00212
324.1 60.1 0.03 100 0.03 100   12.9 -32.728    9.30   14.4 136 16
T 15 03 01 00272
324.3 60.5 0.03 100 0.03 100   13.8 -32.767    7.20   14.3 137 16
T 15 03 01 00332
324.5 61.0 0.03 100 0.02 100   13.9 -32.794    8.67   14.3 138 16
T 15 03 01 00392
324.8 61.4 0.03 100 0.03 100   13.5 -32.843    8.23   14.3 139 16
T 15 03 01 00452
325.0 61.9 0.03 100 0.02 100   13.2 -32.863    8.30   14.2 140 16
```

Figure 9: Example of a scintillation (*.scn) data file for one receiver-satellite pair.

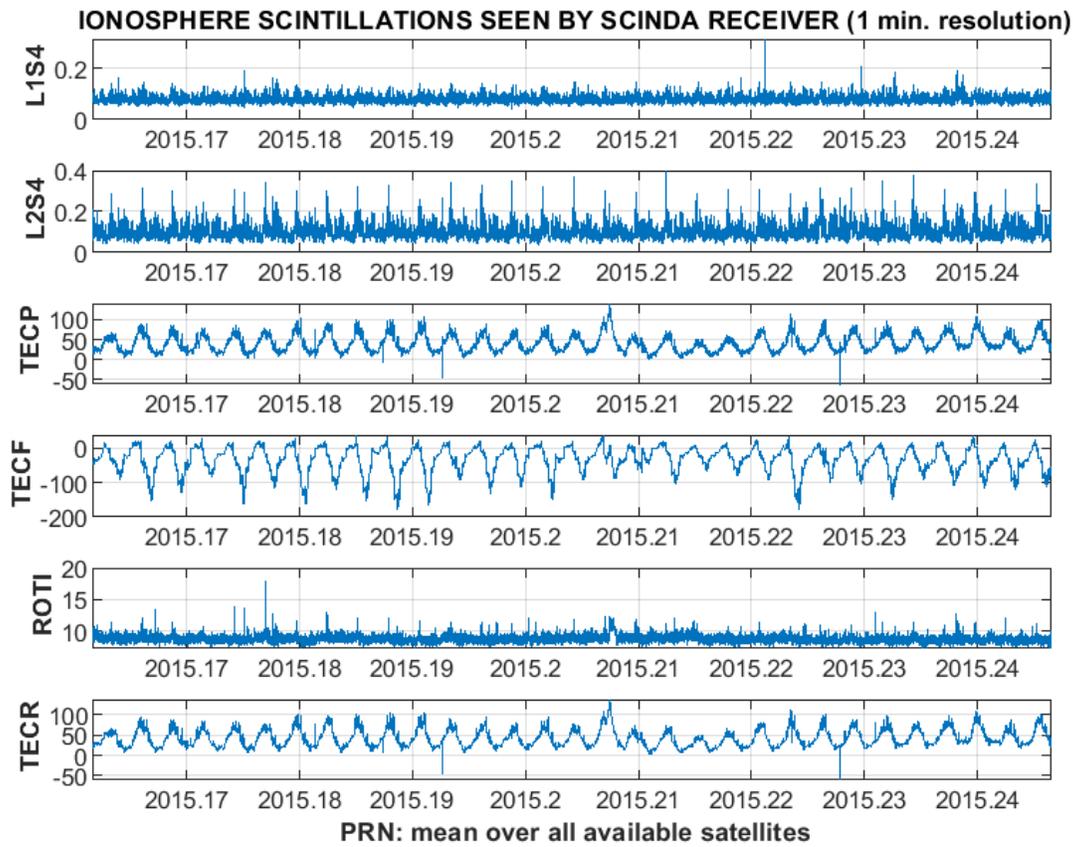

Figure 10: Example of 1-minute resolution plots for an average over all available satellites.

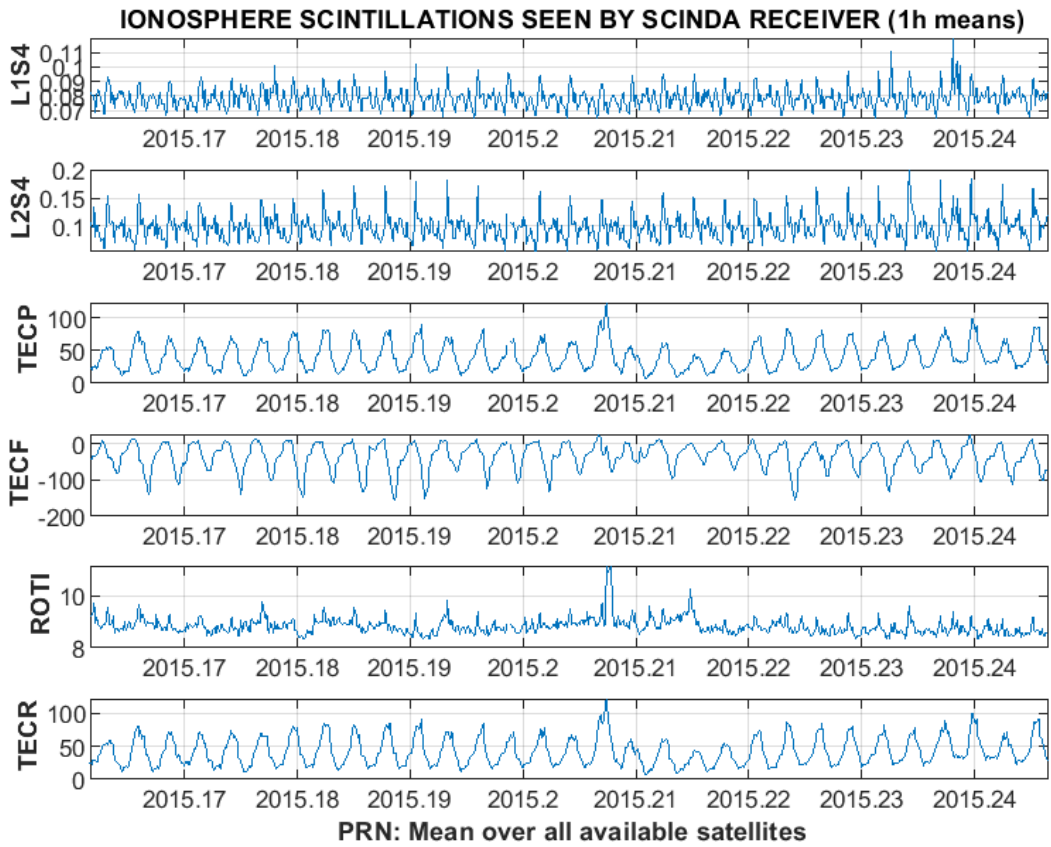

Figure 11: Example of 1-hour resolution plots for an average over all available satellites.

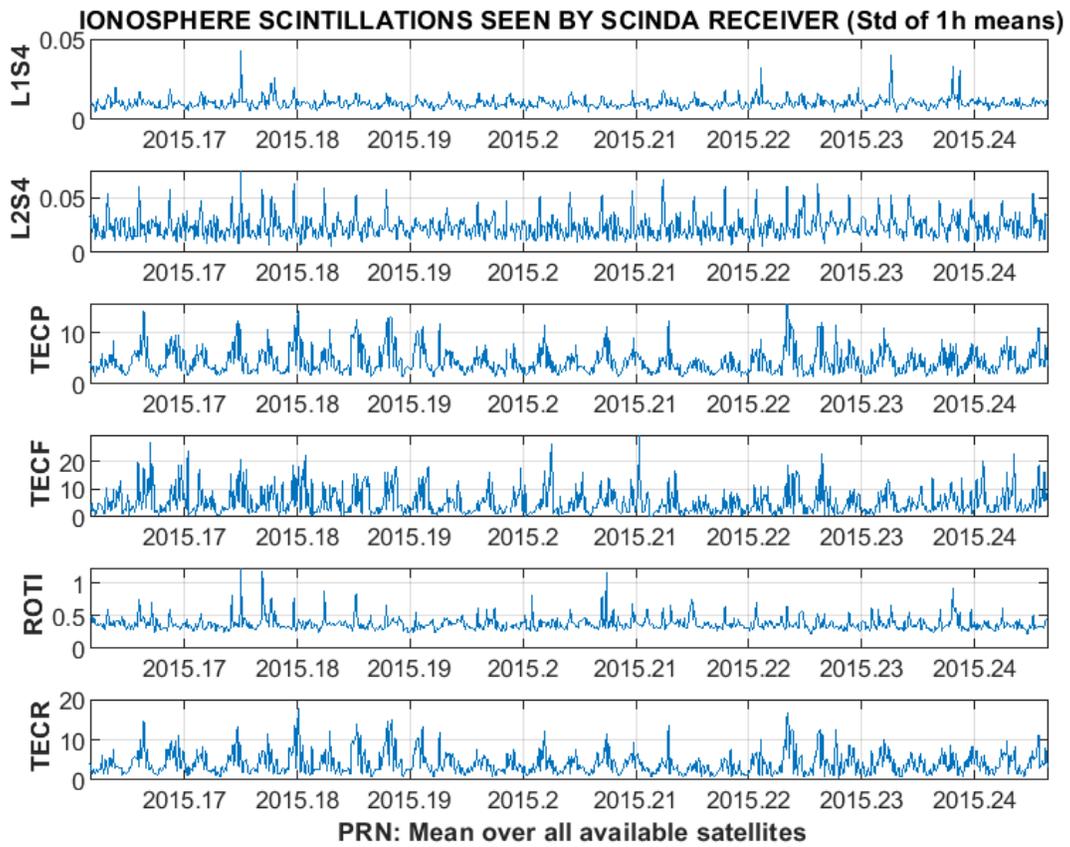

Figure 12: Example of standard deviation plots for 1-hour means calculated over an average over all available satellites.

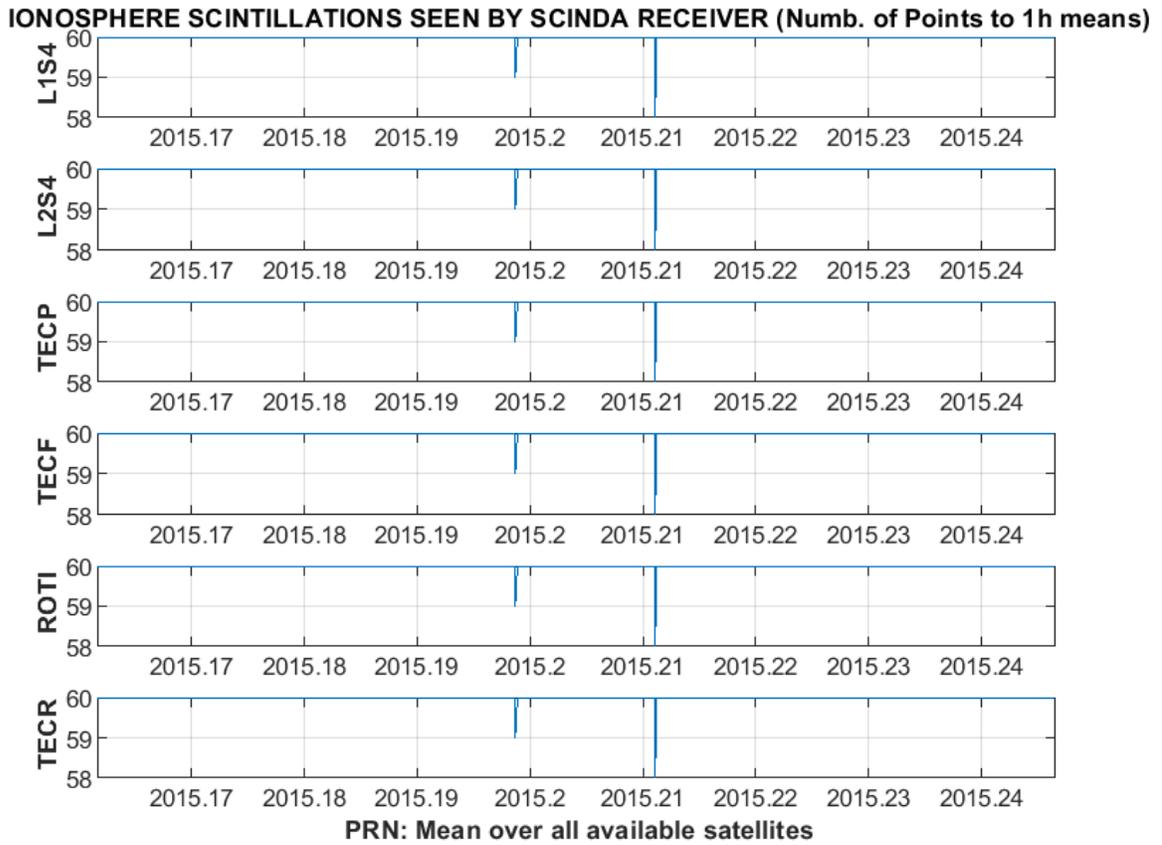

Figure 13: Example of number of successful observations plots for 1-hour means calculated over an average over all available satellites.

**Acknowledgments**

The authors thank Rui Fernandes from SEGAL for the access the SCINDA equipment and the managers of the SCINDA system from Boston College for their help. The SWAIR project founded by the ARTES IAP DEMOSTRATION PROJECTS and CITEUC is founded by FCT, FEDER, COMPETE2020: UID/MULTI/00611/2019; POCI-01-0145-FEDER-006922. The anonymous reviewers are thanked for their help to improve the manuscript.

**Competing Interests**

The authors declare that they have no known competing financial interests or personal relationships which have, or could be perceived to have, influenced the work reported in this article.